\documentstyle[prd,aps]{revtex}
\begin{document}
\title{Sources for the  Majumdar-Papapetrou Space-times}
\author{Metin G{\" u}rses }
\address{ Department of Mathematics, Faculty of Science \\
 Bilkent University, 06533 Ankara - Turkey\\
e-mail: gurses@fen.bilkent.edu.tr
\\[1.5ex]
\begin{minipage}{14cm}\rm\quad
Einstein's  field equations  are  solved  exactly  for static
charged  dust  distributions. These solutions  generalize the
Majumdar-Papapetrou  metrics. Maxwell's equations lead to the
equality of charge and mass densities of the dust distrubition.
Einstein's equations reduce to a nonlinear version of Poisson's
equation.
\\[1ex]
\end{minipage}
}
\maketitle
\def\baselinestretch{1.5}
\section*{Introduction}

Majumdar-Papapetrou (MP) metrics \cite{maj}-\cite{pap} describe the
exterior gravitational field of static objects (discrete point sources,
line charges, point dipoles, charged dusts, etc.,). The line element
of the corresponding metric is given by

\begin{equation}
ds^2=-f\,dt^2+f^{-1}\, \delta_{i\,j}\,dx^{i}\,dx^{j}
\label{le1}
\end{equation}

\noindent
where $f$ is a function of $x^{i}$ and the electromagnetic potential four
vector is given by $A_{u}=(A_{0},0,0,0)$. Letting $f= {1 \over \lambda^2}$
and $A_{0}= {k \over \lambda}$ with $k^2= 1$ one then reduce
the electrovacuum field equations to the Laplace equation $ \nabla ^2
\lambda=0$. The solution of the Laplace equation corresponding to discrete
point sources is given by

\begin{eqnarray}
\lambda= 1+\displaystyle \sum_{i=1}^{N} {m_{i} \over r_{i}},\\
r_{i}=[(x-x_{i})^2+(y-y_{i})^2+(z-z_{i})^2]^{1 \over 2}
\end{eqnarray}

\noindent
In this solution the charge and mass of each point source are
equal, $e_{i}= m_{i}$. These metrics describe
multiple extreme black-hole solutions of Einstein's
theory of gravity in a conformo-static space-time \cite{hah}, \cite{ksm}.
They are also exact solutions of the low energy limit of string theory
\cite{kal} which respect to supersymmetry \cite{gh}-\cite{tod}. There are
also other possible solutions of the Laplace equation but the space-time
geometries corresponding to these solutions have naked singularities
\cite{hah}. To this end it is important to find interior solutions to
remove such singularities.

An extension of the MP metrics
has been studied long time ago by Das \cite{das}. He considered
the charged dust solutions of the static Einstein field equations.
His interest was to find the conditions for an extreme case.
He showed (in his notation) that the $g_{44}$ component of the metric and
$A_{4}$ component of the vector potential are related, $g_{44}=a+b\,A_{4}
+4\pi\,A_{4}^2$, in order for the charge $\rho_{e}$ and mass $\rho$
densities to be equal ,$\rho_{e}=\pm \rho$. He did not investigate the rest
of the field equations.

\section*{Charged Dust Clouds}

In this work we consider the same problem as Das, a generalization
of the MP metrics to the case of a continuous charged dust distribution
(a charged perfect fluid with zero pressure). The energy momentum tensor is
the sum of the energy momentum tensors of the Maxwell field and a dust
distribution. We investigate the complete Einstein field equations for
charged dust distributions in a conformostatic space-time. We first show
that Einstein field equations reduce to a nonlinear potential equation.
We then show that the charge and mass densities are directly related ,
$\rho_{e}=\pm \rho$.. This relation is not an assumption but arises as a
result of the field equations.

Let $M$ be a four dimensional spacetime with the line element (\ref{le1}).
Here Latin letters represent the space indices and $\delta_{ij}$ is the
three dimensional Kronecker delta. In this work we shall use the same
convention as in \cite{hwk}. The only difference is that we use Greek
letters for four dimensional indices. Here $M$ is static. It is useful
to write the metric tensor in a more elegant form. It is given by

\begin{equation}
g_{\mu\, \nu}=f^{-1}\,\eta_{1\, \mu\, \nu}-u_{\mu}\,u_{\nu} \label{mt1}
\end{equation}

\noindent
where $\eta_{1\, \mu\, \nu}=$ \,diag $(0,1,1,1)$ and $u_{\mu}=\sqrt{f}\,
\delta_{\mu}^{0}$. The inverse metric is given by

\begin{equation}
g^{\mu \, \nu}= f\, \eta_{2\,}^{ \mu \, \nu}-f\, \sqrt{f}\,
u^{\mu}\, u^{\nu} \label{mt2}
\end{equation}

\noindent
where $\eta_{2\, \mu\, \nu}=$ \,diag $(0,1,1,1)$ ,
$u^{\mu}=g^{\mu \, \nu}\,u_{\nu}=-{1 \over \sqrt{f}}\, \delta^{\mu}_{0}$.
Here $u^{\mu}$ is a time-like four vector, $u^{\mu}\,u_{\mu}=-1. $

The Ricci scalar and Einstein tensor components corresponding to the conformo
-static metric (\ref{le1}) are given by

\begin{eqnarray}
R={1 \over 2f}\,(2f\, \nabla^2\,f -3\,(\nabla\,f)^2)\\
G_{00}=f\, \nabla^2\,f -{5 \over 4}\, (\nabla f)^2 ,~~~G_{0i}=0\\
G_{ij}=-{1 \over 2f^2}\, \partial_{i}\,f\, \partial_{j}\,f +
{(\nabla\,f)^2 \over 4f^2}\, \delta_{ij}
\end{eqnarray}

\noindent
where $\nabla^2$ denotes the three three dimensional Laplace operator in 
Cartesian flat coordinates.
The Maxwell antisymmetric tensor  and the corresponding energy momentum
tensor are respectively given by

\begin{eqnarray}
F_{\mu\, \nu}= \nabla_{\nu}\,A_{\mu}-\nabla_{\mu}\,A_{\nu}\\
M_{\mu \, \nu}={1 \over 4 \pi}\,(F_{\mu\, \alpha}\, F_{\nu}^{\alpha}-
{1 \over 4}\,F^2\,g_{\mu \, \nu})
\end{eqnarray}

\noindent
where $F^2= F^{\mu\, \nu}\,F_{\mu \, \nu}$. The current vector $j^{\mu}$
is defined as

\begin{equation}
\nabla_{\nu}\,F^{\mu\, \nu}=4\pi\,j^{\mu}
\end{equation}

\noindent
Here we should assume $A_{i}=0$. Then

\begin{eqnarray}
M_{00}={f \over 8\pi}\, (\nabla A_{0})^2 , ~~M_{0i}=0~~,~~~
M_{ij}=-{1 \over 4\pi\,f}\, \partial_{i}\,A_{0}\, \partial_{j}\, A_{0}+
{1 \over 8\pi\,f}\, (\nabla A_{0})^2\, \delta_{ij}
\end{eqnarray}

\noindent
and

\begin{eqnarray}
4\pi\,j^{0}=f\, \nabla\,(f^{-1}\, \nabla \,A_{0}), ~~~j^{i}=0.
\end{eqnarray}

\noindent
The Einstein field equations for a charged dust distribution are given by

\begin{equation}
G_{\mu\, \nu}=8\, \pi\,T_{\mu\, \nu}=8 \pi\,M_{\mu \, \nu}+
(8 \, \pi\, \rho)\, u_{\mu}\,u_{\nu}
\end{equation}

\noindent
where $\rho$ is the energy density of the charged dust distribution and the
four velocity of the dust is the same vector $u^{\mu}$ appearing in the
metric tensor. We find that $j^{\mu}=-\rho_{e}\, u^{\mu}$, where $\rho_{e}$

\begin{equation}
\rho_{e}={1 \over 4\pi}\,f^{3/2}\, \nabla \,(f^{-1}\, \nabla\, A_{0})
\end{equation}

\noindent
is the charge density of the dust distribution. Let $\lambda$ be a real
function depending on the space coordinates $x^{i}$. Einstein's equations $
G_{ij}=8\pi \,T_{ij}$  forces us to choose $A_{0}=k\,\sqrt{f}$ or

\begin{equation}
f={1 \over \lambda^2}, ~~A_{0}={k \over \lambda}
\end{equation}

\noindent
where $k=\pm 1$. Then the remaining field equation $G_{00}=8\pi\,T_{00}$ 
reduce simply to the following
equations.

\begin{eqnarray}
\nabla^2\, \lambda +4\, \pi \, \rho\, \lambda^3=0 \label{eq2} \\
\rho_{e}=k\, \rho
\end{eqnarray}

These equations represent the Einstein and Maxwell
equations respectively. In particular the first equation (\ref{eq2})
is a generalization of the Poisson's potential equation in
Newtonian gravity. In the Newtonian approximation $\lambda=1+V$,
Eq.(\ref{eq2}) reduces to the Poisson equation, $\nabla^2\, V+4\pi\, \rho=0$.
Hence for any physical mass density $\rho$ of the dust distribution
we solve the equation (\ref{eq2}) to find the function
$\lambda$. This determines the space-time metric completely.
As an example for a constant mass density $\rho=\rho_{0} >0$ we find that

\begin{equation}
\lambda=\displaystyle {a \over 2\, \sqrt{\pi\, \rho_{0}}}\,cn (l_{i}\, x^{i})
\end{equation}

\noindent
Here $l_{i}$ is a constant three vector ,$a^2=l_{i}l^{i}$ and $cn$ is one of
the Jacobi elliptic function with modulus square equals  ${1 \over 2}$.
This is a model universe which is filled by a (extreme) charged dust
with a constant mass density.

\section*{Interior Solutions}

In an asymptotically flat space-time , the function $\lambda$
asymptotically obeys the boundary condition $\lambda \rightarrow 1$.
In this case we can establish the equality of mass and charge
$e=\pm m_{0}$, where $m_{0}= \int \rho\, \sqrt{-g}\,d^3 x$. For physical
considerations our extended MP space-times
may be divided into inner and outer regions.
The inner and outer regions are defined as the regions where
$\rho_{i} >0$ and $\rho=0$ respectively. Here $i=1,2,...,N$, where
$N$ represents the number of regions. The gravitational fields of
the outer regions are described by any solution of the Laplace
equation $\nabla^2\, \lambda=0$ , for instance by the MP metrics.
As an example the extreme Reissner-Nordstr{\" o}m (RN) metric
(for $r >R_{0}$), $\lambda = 1+\displaystyle {m_{0} \over r}$ may be
matched to a metric with

\begin{equation}
\lambda=a\, \displaystyle { \sin (b\,r) \over r},\,\, r < R_{0}
\end{equation}

\noindent
describing the gravitational field of an inner region filled by
a spherically symmetric charged dust distribution with
a mass density

\begin{equation}
\rho=\displaystyle \rho(0)\,[{b\, r \over \sin (b\,r)}]^{2}
\label{rho}
\end{equation}

\noindent
Here $\rho(0)={1 \over 4\, \pi\, a^2}$, $r^2=x_{i}\,x^{i}$,
$a$ and $b$ are constants to be determined
in terms of the radius $R_{0}$ of the boundary and total mass $m_{0}$
(or in terms of $\rho(0)$). They are given by

\begin{equation}
b\,R_{0}=\displaystyle \sqrt{ {3m_{0} \over R_{0}}}~,~~a\, \sin b\,R_{0}=
R_{0}+m_{0}
\end{equation}

\noindent
In this way one may eliminate
the singularities of the outer solutions by matching them to an inner
solution with a physical mass density.

\noindent
We can extend the above solution to $N$ charged dust spheres by letting
the mass density $\rho= {b^{2} \over  4 \pi\,\lambda^2}$  where we have the 
complete solution of (\ref{eq2}).

\begin{equation}
\lambda=\displaystyle \sum_{l,m}\,a_{l,m}\,
j_{l}(b\,r)\,Y_{l,m}(\theta, \phi)
\end{equation}

\noindent
where $j_{l}(b\,r)$ are the spherical Bessel functions which are
given by

\begin{equation}
j_{l}(x)=\displaystyle (-x)^l\,\big ( {1 \over x}\, {d \over dx} \big )^l\,
\big ( {\sin x \over x} \big )
\end{equation}

\noindent
and $Y_{l,m}$ are the spherical harmonics. The constants
$a_{l,m}$ are determined when this solution is matched to an outer solution
with $\nabla^2 \lambda=0$. The interior solution given above for
the extreme RN metric ($N=1$) with density (\ref{rho}) correspond to $l=0$.

\section*{Concluding Remarks}

In this work we have solved the Einstein field equations in a
conformo-static space-time for a charged dust distribution. We
reduced the problem to a nonlinear Poisson type of potential equation
(\ref{eq2}). Any physically reasonable solution of this equation
gives an interior solution to an exterior MP metric with naked
singularities. In this way we remove the naked singularities of the MP
metrics by matching them to the metrics of extremely charged dust
distributions.
We have given some explicit exact solutions corresponding to some specific
mass densities. In particular we have given an interior solution of the
extreme RN metric.

In reality such objects may or may not exist. If they exist, such charged
dust clouds can not be detected directly , because they may transmit
light rays. On the other hand they may be observed through the bending
of light rays passing very close to them. The measure of deflection
angles will be of the same order of magnitude of the
deflection angle of the null geodesics in the Schwarzschild geometry  with
the same total mass $m_{0}$. The reason for this becomes obvious when we write
the RN metric in the Schwarzschild coordinates

\begin{equation}
ds^2=\displaystyle -(1- {m_{0} \over r})^2\, dt^2+ {dr ^2
\over( 1-{m_{0} \over r})^2}+r^2\,d \Omega^2.
\end{equation}

\noindent
For a charged dust cloud with  ${m_{0} \over R_{0}} <1$ and $r>R_{0}$
we have $(1-{m_{0} \over r})^2 \rightarrow (1- {2m_{0} \over r})$. Thus
in the neighborhood of the charged dust cloud, the exterior
metric becomes closer to the Schwarzschild metric. We then expect the same
effects as in the Schwarzschild geometry on the test particles in the
neighborhood of extremely charged dust clouds.
The measure of the deflection angle may be much larger if these light rays
are also transmitted through the dust clouds.

This work is partially supported by Turkish Academy of Sciences
and also by the Scientific and Technical Research Council of Turkey.

\end{document}